# Freezing dynamics of the ferrofluid droplet in a uniform magnetic field using the lattice Boltzmann flux solver

**Jinxiang Zhou[1], Liming Yang[1, 2, 3, *], Yaping Wang[1], Jie Wu[1], Xiaodong Niu[4]**

[1]Department of Aerodynamics, College of Aerospace Engineering, Nanjing University of Aeronautics and Astronautics, Nanjing 210016, China

[2]State Key Laboratory of Mechanics and Control for Aerospace Structures, Nanjing University of Aeronautics and Astronautics, Nanjing 210016, China

[3]MIIT Key Laboratory of Unsteady Aerodynamics and Flow Control, Nanjing University of Aeronautics and Astronautics, Nanjing 210016, China

[4]College of Engineering, Shantou University, 243 Daxue Road, Shantou 515063, Guangdong, China

**Abstract**

In this study, an enthalpy-based lattice Boltzmann flux solver is developed to simulate the freezing dynamics of a ferrofluid droplet under a uniform magnetic field. The accuracy and robustness of the solver are first validated through three benchmark tests: conductive freezing, static droplet freezing, and ferrofluid droplet deformation. The solver is then employed to investigate the influence of a uniform magnetic field on the freezing behavior of ferrofluid droplets, with particular emphasis on the overall freezing process, heat transfer characteristics, and freezing duration. The results reveal that the uniform magnetic field affects the freezing dynamics primarily by altering the droplet morphology. Under a vertically oriented magnetic field, the droplet elongates along the field direction, which increases the thermal resistance and consequently prolongs the freezing time. Conversely, a horizontally uniform magnetic field flattens the droplet, reducing the thermal resistance and thus shortening the freezing time. These findings provide new physical insight into magnetic-field-induced modulation of the freezing process in ferrofluid systems.

---

*Corresponding author, E-mail: lmyang@nuaa.edu.cn.

1. **Introduction**

Droplet freezing is a common phenomenon in both industrial and natural contexts. In aerospace engineering, ice accumulation on aircraft propellers can lead to power loss and pose serious flight safety risks [1]-[4]. Similarly, frozen power transmission lines suffer from excessive load and collapse hazards [5], while ice-covered photovoltaic panels experience severe reductions in power output [6]. Owing to these widespread impacts, developing efficient methods to prevent or mitigate droplet freezing on supercooled surfaces has become a critical research focus, drawing increasing attention in recent years [7]-[9].

Current anti-icing and de-icing techniques can be broadly categorized as passive or active. Passive approaches aim to delay ice nucleation by modifying surface properties. For example, Qiao et al. [10] fabricated a photothermal superhydrophobic melamine sponge surface with outstanding chemical stability, thermal durability, and reusability for de-icing applications. Oberli et al. [11] reported that superhydrophobic surfaces not only repel water effectively but also reduce ice adhesion and delay freezing. However, such surfaces often degrade under thermal cycling or prolonged wetting, reducing their anti-icing effectiveness. Moreover, passive strategies face limitations such as the high cost of nanomaterials and the increased thermal resistance introduced by coatings, which hinders convective heat transfer and de-icing efficiency. Therefore, more effective and controllable active approaches are increasingly being explored.

Active de-icing methods employ external fields, such as ultrasonic waves [12], electric fields [13]-[15], and magnetic fields [16]-[18], to actively regulate the freezing process. For instance, Gao et al. [12] found that ultrasonic waves accelerate mass transfer, leading to faster cooling, with droplet temperatures dropping by 2.0°C-2.5°C compared to the non-ultrasonic case. Deng et al. [15] reported that an applied electric field reduces nucleation activation energy and increases the freezing temperature, though it extends the time required for complete solidification. Inaba et al. [19] demonstrated that the variation in freezing point temperature under a magnetic field follows a quadratic dependence on the field strength.

Numerous experimental studies have examined the freezing behavior of water droplets [20]-[24]. For example, Chang et al. [20] analyzed the effects of impact velocity and substrate temperature on droplet morphology, including spreading behavior, height ratio, and freezing duration across various substrate materials. Zhang et al. [22] examined droplet impact and freezing on cold concave surfaces, reporting ring-shaped residues after recoating and noting that surface curvature had little influence on axial spreading. Jin et al. [24] studied continuous droplet freezing on ice surfaces and found that both the ice surface temperature and the droplet release height significantly affected not only the freezing time but also the final contact diameter.

However, most experimental studies focus on the macroscopic aspects of freezing while overlooking microscopic phenomena such as interfacial dynamics and thermal transport processes. With advances in computational power, numerical simulations have become indispensable tools for studying droplet freezing [25]-[30]. Lyu et al. [29] developed a hybrid volume-of-fluid immersed-boundary (VOF-IB) method that accounts for volume expansion effects during droplet freezing, and validated their model against experimental and theoretical results. Thirumalaisamy et al. [31] proposed an enthalpy method to simulate freezing and melting with variable thermophysical properties, focusing particularly on liquid pools containing bubbles. Huang et al. [32] introduced an enthalpy lattice Boltzmann method (LBM) for droplet and bubble freezing, while Zhang et al. [33] proposed an axisymmetric lattice Boltzmann model that incorporates droplet volume expansion. Using this model, they analyzed the effects of contact angle and volume expansion on freezing time and the formation of the conical tip at the top of the frozen droplet.

Ferrofluids, which are stable colloidal suspensions of magnetic nanoparticles in an organic carrier fluid, exhibit unique gel-like properties that make them suitable for a wide range of applications, including lubrication [34, 35], sealing [36, 37], thermal management [38, 39], and microfluidics [40, 41]. Although extensive studies have been conducted on water droplet freezing on supercooled surfaces, research on ferrofluid freezing remains comparatively limited, particularly numerical investigations. Experimental efforts have explored certain aspects of ferrofluid solidification [42]-[45]. For instance, Morales et al. [43] examined the freezing behavior of two $Fe_3O_4$-based ferrofluids with 14 nm nanoparticles dispersed in n-hexane and dodecane, finding markedly different freezing points (below 178 K for n-hexane and above 264 K for dodecane). Yang et al. [45] investigated the magnetization and magnetic viscosity of aqueous ferrofluids before and after solidification, revealing enhanced saturation magnetization and reduced magnetic viscosity in the frozen state under increasing magnetic field strength. Application-oriented studies include Irajizad et al. [44], who developed a ferrofluid-based surface achieving a $10^3$-fold increase in freezing delay with ultralow adhesion, and Zhang et al. [42], who proposed a predictive model for ferrofluid droplet deformation in a non-uniform magnetic field and analyzed its macroscopic freezing behavior. However, to the best of the authors' knowledge, only Fang et al. [46] have numerically investigated the effect of a non-uniform magnetic field on the freezing time of a ferrofluid droplet.

In this study, an enthalpy-based lattice Boltzmann flux solver is developed to investigate the freezing dynamics of ferrofluid droplets in a uniform magnetic field. The research primarily focuses on the overall freezing process, heat transfer characteristics, and freezing time of the ferrofluid droplet. The structure of this paper is organized as follows: Section 2 introduces the governing equations; Section 3 details the numerical methodology; Section 4 validates the model against experimental and theoretical benchmarks; Section 5 presents and analyzes the simulation results for uniform magnetic-field-induced freezing dynamics; and Section 6 concludes the study.

## 2. Governing equation

The freezing process of a ferrofluid droplet is a multi-physics coupling problem that involves interactions among the magnetic, thermal, flow, and phase fields.

First of all, the Maxwell equations for a static uniform magnetic field can be expressed as [47]:

$$\begin{cases} \nabla \cdot \mathbf{B} = 0 \\ \nabla \times \mathbf{H} = 0 \end{cases}, \quad (1)$$

where **H** is the magnetic field intensity and **B** is the magnetic flux density. The magnetic flux density can be calculated as:

$$\mathbf{B} = \mu_0 (\mathbf{H} + \mathbf{M}) = \mu_0 (1 + \chi) \mathbf{H} = \mu_m \mathbf{H}, \quad (2)$$

where **M** is the magnetization, $\chi$ is the magnetic susceptibility, $\mu_0$ is the vacuum permeability, and $\mu_m$ is the magnetic permeability of the medium. By introducing a magnetic scalar potential $\phi_m$, the magnetic intensity can be described as:

$$\mathbf{H} = -\nabla \phi_m. \quad (3)$$

Substituting Eqs. (2) and (3) into the first equation of Eq. (1), the magnetic potential equation can be obtained:

$$\nabla \cdot (\mu_m \nabla \phi_m) = 0. \quad (4)$$

The motion of the ferrofluid droplet in the uniform magnetic field is influenced by magnetic stresses acting at the fluid interface, which are represented by the Maxwell stress tensor[48]:

$$\boldsymbol{\tau}_m = -\frac{\mu_0}{2} |\mathbf{H}|^2 \mathbf{I} + \mathbf{H}\mathbf{B}, \quad (5)$$

where $\upsilon$ is the kinematic viscosity and **I** is the identity tensor. Then, the magnetic force at the fluid interface can be calculated as:

$$\begin{aligned}
\mathbf{F}_{mg} &= \nabla \cdot \boldsymbol{\tau}_m = -\frac{\mu_0}{2}\nabla\left(|\mathbf{H}|^2\right) + \left(\nabla\cdot\mu_m\mathbf{H}\right)\mathbf{H} + \left(\mu_m\mathbf{H}\cdot\nabla\right)\mathbf{H} \\
&= -\frac{\mu_0}{2}\nabla\left(|\mathbf{H}|^2\right) + \left(\mu_m\mathbf{H}\cdot\nabla\right)\mathbf{H} \\
&= -\frac{\mu_0}{2}\nabla\left(|\mathbf{H}|^2\right) + \mu_m\left[\frac{1}{2}\nabla(\mathbf{H}\cdot\mathbf{H}) - \mathbf{H}\times(\nabla\times\mathbf{H})\right] \\
&= \frac{\mu_m - \mu_0}{2}\nabla\left(|\mathbf{H}|^2\right) \\
&= \frac{\mu_0\chi}{2}\nabla\left(|\mathbf{H}|^2\right)
\end{aligned} \quad (6)$$

Secondly, the evolution of the solid-liquid interface during the solidification process is governed by the total enthalpy equation [32]:

$$\frac{\partial H_e}{\partial t} + \nabla\cdot(C_p T\mathbf{u}) = \nabla\cdot\left(\frac{\lambda}{\rho}\nabla T\right) + C_p T\nabla\cdot\mathbf{u}, \quad (7)$$

where $H_e$, $T$, $\rho$, $\mathbf{u}$, $\lambda$ and $C_p$ are the total enthalpy, temperature, density, velocity, thermal conductivity, and specific heat at constant pressure, respectively. $H_e$ can be calculated as $H_e = h + LF_l = C_p T + LF_l$, where $h$, $L$, and $F_l$ are the sensible enthalpy, latent heat, and liquid fraction, respectively. The liquid fraction and temperature can be determined from the total enthalpy $H_e$ by [51]:

$$F_l = \begin{cases} 0 & H_e < H_{es} \\ \dfrac{H_e - H_{es}}{H_{el} - H_{es}} & H_{es} \leq H_e \leq H_{el} \\ 1 & H_{ee} > H_{el} \end{cases}$$

$$T = \begin{cases} \dfrac{H_e}{C_p} & H_e < H_{es} \\ T_s + \dfrac{H_e - H_{es}}{H_{el} - H_{es}}(T_l - T_s) & H_{es} \leq H_e \leq H_{el} \\ T_l + \dfrac{H_e - H_{el}}{C_p} & H_e > H_{el} \end{cases} \quad (8)$$

where $H_{es} = C_{p,s} T_s$ and $H_{el} = C_{p,l} T_l + L$ are the total enthalpy at solidus and liquidus temperatures, respectively.

Thirdly, the flow and phase fields are governed by the incompressible Navier-Stokes (NS) and Cahn-Hilliard (CH) equations with the volume change term [32]:

$$\begin{cases} \dfrac{\partial p}{\partial t} + \rho c_s^2 \nabla \cdot \mathbf{u} = \rho c_s^2 \left(1 - \dfrac{\rho_s}{\rho_l}\right) \dfrac{\partial (F_s)}{\partial t} \\ \dfrac{\partial \rho \mathbf{u}}{\partial t} + \nabla \cdot (\rho \mathbf{u}\mathbf{u}) = -\nabla p + \nabla \cdot \left[\mu\left(\nabla \mathbf{u} + (\nabla \mathbf{u})^T\right)\right] + \mathbf{F}_{surf} + \mathbf{F}_g + \mathbf{F}_{mg} + \mathbf{F}_{IBM} \end{cases} \quad (9)$$

$$\frac{\partial C}{\partial t} + \nabla \cdot (C\mathbf{u}) = \nabla \cdot (M_c \nabla \mu_c) + C\left(1 - \frac{\rho_s}{\rho_l}\right)\frac{\partial F_s}{\partial t}, \quad (10)$$

where $p$, $t$, $\mu$, $C$ are the pressure, time, dynamic viscosity, and order parameter, respectively. $M_c$ and $\mu_c$ are the mobility and chemical potential. $\rho_s$ and $\rho_l$ are the densities of solid and liquid. $\mathbf{F}_{surf} = -C\nabla\mu_c$ is the surface tension force, $\mathbf{F}_g = \rho g$ is the gravity, and $g$ is the gravitational acceleration. $\mathbf{F}_{mg}$ is the electric force. $\mathbf{F}_{IBM} = -\rho \mathbf{u}(\partial F_s/\partial t)$ is the body force [29]. $F_s$ is the solid fraction, which is related to the liquid fraction $F_l$.

The chemical potential is given by $\mu_c = 2\beta C(C-1)(2C-1) + \kappa \nabla^2 C$, with $\beta = 12\sigma/\xi$ and $\kappa = 3\xi\sigma/2$, where $\sigma$ is the interfacial tension coefficient and $\xi$ is the interface thickness. Additionally, the thermophysical and magnetic properties of the ferrofluid are determined by:

$$\gamma = F_s \gamma_s + (1 - F_s) C \gamma_l + (1 - F_s)(1 - C)\gamma_g, \quad (11)$$

where $\gamma$ represents any relevant physical property, and subscripts $s$ and $l$ denote the solid and liquid phases, respectively.

In summary, Eqs. (4), (7), (9), and (10) together constitute the governing equations for the freezing process of ferrofluid droplets.

## 3. Numerical method
### 3.1 Lattice Boltzmann flux solver for the magnetic field

The single relaxation parameter lattice Boltzmann equation(LBE) corresponding to Eq. (4) can be expressed as:

$$m_\alpha(\mathbf{x}+\mathbf{e}_\alpha \delta_t, t+\delta_t) - m_\alpha(\mathbf{x},t) = \frac{m_\alpha^{eq}(\mathbf{x},t) - m_\alpha(\mathbf{x},t)}{\tau_m}, \tag{12}$$

where $m_\alpha$, $m_\alpha^{eq}$, and $\tau_m = 0.5 + (\mu_m D)/(c_s^2 \delta_t)$ represent the distribution function of the magnetic potential, its equilibrium state, and the single relaxation parameter for the magnetic field, respectively. $\delta_t$ is the streaming time step size and $\mathbf{e}_\alpha$ is the lattice speed. *D* is a free parameter corresponding to the evolution speed. The equilibrium state of the magnetic potential is defined as:

$$m_\alpha^{eq}(\mathbf{x},t) = \omega_\alpha \phi_m, \quad \alpha = 0-8 \tag{13}$$

In this study, the D2Q9 model [49] is adopted, as shown in Fig. 1. The corresponding weight coefficients are given by $\omega_0 = 4/9$, $\omega_{1-4} = 4/9$ and $\omega_{5-8} = 4/9$, and the lattice speed is defined as:

$$\mathbf{e}_\alpha = \begin{cases} (0,0), & \alpha = 0 \\ c\left[\cos\left(\frac{(\alpha-1)\pi}{2}\right), \sin\left(\frac{(\alpha-1)\pi}{2}\right)\right], & \alpha = 1-4 \\ \sqrt{2}c\left[\cos\left(\frac{(\alpha-1)\pi}{2}\right), \sin\left(\frac{(\alpha-1)\pi}{2}\right)\right], & \alpha = 5-8 \end{cases} \tag{14}$$

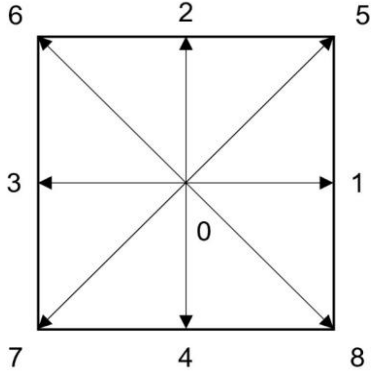

Fig. 1. Schematic diagram of the D2Q9 model.

Using the Chapman-Enskog expansion analysis and considering that $\sum_\alpha m_\alpha^{eq} = \phi_m$ and $\sum_\alpha \mathbf{e}_\alpha m_\alpha^{eq} = 0$, the macroscopic equation for the magnetic field can be recovered from Eq. (12) (details provided in Ref. [50]):

$$\frac{\partial \phi_m}{\partial t} + \nabla \cdot \left[ \left( \sum_\alpha \mathbf{e}_\alpha m_\alpha^{eq} \right) + \left( 1 - \frac{1}{2\tau_m} \right) \left( \sum_\alpha \mathbf{e}_\alpha m_\alpha^{neq} \right) \right] = 0, \quad (15)$$

where $m_\alpha^{neq}(\mathbf{x},t) = -\tau_m \left[ m_\alpha^{eq}(\mathbf{x},t) - m_\alpha^{eq}(\mathbf{x}-\mathbf{e}_\alpha \delta_t, t-\delta_t) \right]$. Define $m_\alpha^*$ as:

$$m_\alpha^* = \left[ m_\alpha^{eq} + \left( 1 - \frac{1}{2\tau_m} \right) m_\alpha^{neq} \right]. \quad (16)$$

Then, Eq. (15) can be simplified as:

$$\frac{\partial \phi_m}{\partial t} + \nabla \cdot \sum_\alpha \mathbf{e}_\alpha m_\alpha^* = 0. \quad (17)$$

Using the definitions of equilibrium and non-equilibrium terms, the relations between the macroscopic fluxes and the distribution function can be derived as:

$$D\mu_m \nabla \phi_m = \sum_\alpha \mathbf{e}_\alpha m_\alpha^*. \quad (18)$$

### 3.2 Lattice Boltzmann flux solver for the temperature field

The single relaxation parameter LBE corresponding to Eq. (7) can be expressed as:

$$h_\alpha(\mathbf{x}+\mathbf{e}_\alpha \delta_t, t+\delta_t) - h_\alpha(\mathbf{x},t) = \frac{h_\alpha^{eq}(\mathbf{x},t) - h_\alpha(\mathbf{x},t)}{\tau_h}, \quad \alpha = 0-8 \quad (19)$$

where $h_\alpha$, $h_\alpha^{eq}$ and $\tau_h = 0.5 + \lambda/(\rho C_{p,ref} c_s^2 \delta_t)$ represent the distribution function of the total enthalpy, its equilibrium state, and the single relaxation parameter for the temperature field, respectively. The reference specific heat capacity is defined as $C_{p,ref}=(C_{p,s}+C_{p,l})/2$, where $C_{p,s}$ and $C_{p,l}$ denote the specific heats of the solid and liquid phases, respectively. The equilibrium state $h_\alpha^{eq}$ is given by:

$$h_\alpha^{eq} = \begin{cases} H_e - C_{p,ref} T + \omega_\alpha C_p T \left( \dfrac{C_{p,ref}}{C_p} - \dfrac{\mathbf{u}^2}{2c_s^2} \right), & \alpha = 0 \\ \omega_\alpha C_p T \left( \dfrac{C_{p,ref}}{C_p} + \dfrac{\mathbf{e}_\alpha \cdot \mathbf{u}}{c_s^2} + \dfrac{(\mathbf{e}_\alpha \cdot \mathbf{u})^2}{2c_s^4} - \dfrac{\mathbf{u}^2}{2c_s^2} \right), & \alpha = 1-8 \end{cases} \quad (20)$$

where $\omega_\alpha$ and $c_s$ are the weight coefficients and sound speed, respectively. Applying the Chapman-Enskog expansion analysis and considering that $\sum_{\alpha} h_{\alpha}^{eq} = H_e$ and $\sum_{\alpha} \mathbf{e}_{\alpha} h_{\alpha}^{eq} = C_p T \mathbf{u}$, the following equation can be derived from Eq. (19) (details provided in the Appendix):

$$\frac{\partial H_e}{\partial t} + \nabla \cdot \left[ \sum_{\alpha} \mathbf{e}_{\alpha} \left( h_{\alpha}^{eq} + \left(1 - \frac{1}{2\tau_h}\right) h_{\alpha}^{neq} \right) \right] = 0, \quad (21)$$

where $h_{\alpha}^{neq}(\mathbf{x},t) = -\tau_h \left[ h_{\alpha}^{eq}(\mathbf{x},t) - h_{\alpha}^{eq}(\mathbf{x} - \mathbf{e}_{\alpha}\delta_t, t - \delta_t) \right]$. Define $h_{\alpha}^*$ as:

$$h_{\alpha}^* = \left[ h_{\alpha}^{eq} + \left(1 - \frac{1}{2\tau_h}\right) h_{\alpha}^{neq} \right]. \quad (22)$$

Then, Eq. (21) can be simplified as:

$$\frac{\partial H_e}{\partial t} + \nabla \cdot \sum_{\alpha} \mathbf{e}_{\alpha} h_{\alpha}^* = 0. \quad (23)$$

Using the definitions of equilibrium and non-equilibrium terms, the relations between the macroscopic fluxes and the distribution function can be derived as:

$$C_p T \mathbf{u} - \frac{\lambda}{\rho} \nabla(T) = \sum_{\alpha} \mathbf{e}_{\alpha} h_{\alpha}^*. \quad (24)$$

Finally, the macroscopic equation for the temperature field can be expressed as:

$$\frac{\partial H_e}{\partial t} + \nabla \cdot \sum_{\alpha} \mathbf{e}_{\alpha} h_{\alpha}^* - C_p T \nabla \cdot \mathbf{u} = 0. \quad (25)$$

### 3.3 Lattice Boltzmann flux solver for the flow and phase fields

The standard LBEs for the flow and phase fields can be expressed as:

$$f_{\alpha}(\mathbf{x} + \mathbf{e}_{\alpha}\delta_t, t + \delta_t) - f_{\alpha}(\mathbf{x},t) = \frac{f_{\alpha}^{eq}(\mathbf{x},t) - f_{\alpha}(\mathbf{x},t)}{\tau_f}, \alpha = 0-8, \quad (26)$$

$$g_{\alpha}(\mathbf{x} + \mathbf{e}_{\alpha}\delta_t, t + \delta_t) - g_{\alpha}(\mathbf{x},t) = \frac{g_{\alpha}^{eq}(\mathbf{x},t) - g_{\alpha}(\mathbf{x},t)}{\tau_g}, \alpha = 0-8 \quad (27)$$

where $f_{\alpha}$, $f_{\alpha}^{eq}$ and $\tau_f = 0.5 + \mu/(\rho c_s^2 \delta_t)$ represent the distribution function of the flow field, its equilibrium state, and the single relaxation parameter for the flow field, respectively. $g_{\alpha}$, $g_{\alpha}^{eq}$ and $\tau_g$ denote the distribution function of the order parameter, its equilibrium state, and the single relaxation parameter for the phase field, respectively. $\tau_g$ is related to the diffusion parameter $Q_g$ and the mobility $M_c$ as $M_c = (\tau_g - 0.5) Q_g \delta_t$. The equilibrium states $f_\alpha^{eq}$ and $g_\alpha^{eq}$ are given by:

$$f_\alpha^{eq}(\mathbf{x},t) = \omega_\alpha \left[ p + \rho c_s^2 \left( \frac{(\mathbf{e}_\alpha \cdot \mathbf{u})}{c_s^2} + \frac{(\mathbf{e}_\alpha \cdot \mathbf{u})^2}{2c_s^4} - \frac{\mathbf{u}^2}{2c_s^2} \right) \right], \alpha = 0-8 \quad (28)$$

$$g_\alpha^{eq}(\mathbf{x},t) = \begin{cases} C - \dfrac{\mu_c Q_g (1-\omega_0)}{c_s^2}, & \alpha=0 \\ \dfrac{\omega_\alpha (\mu_c Q_g + C \mathbf{e}_\alpha \cdot \mathbf{u})}{c_s^2}, & \alpha=1-8 \end{cases} \quad (29)$$

Using the Chapman-Enskog expansion analysis and introducing a new distribution function $f_\alpha^{eq-m}(\mathbf{x} - \mathbf{e}_\alpha \Delta t, t - \Delta t) = \omega_\alpha \left[ p + \rho(\mathbf{x}, t-\Delta t) \Gamma_\alpha \right]$ for the flow field at the surrounding points of the cell interface, Eqs. (26) and (27) can be reformulated as [52-54] (details are provided in Refs. [52-54]):

$$\begin{cases} \dfrac{\partial p}{\partial t} + \nabla \cdot \left( \sum_\alpha \mathbf{e}_\alpha f_\alpha^{eq} \right) = 0 \\ \dfrac{\partial \rho c_s^2 \mathbf{u}}{\partial t} + \nabla \cdot \sum_\alpha \mathbf{e}_\alpha \mathbf{e}_\alpha \left[ f_\alpha^{eq} - \left(1 - \dfrac{1}{2\tau_f}\right) f_\alpha^{neq-m} \right] = 0 \end{cases}, \quad (30)$$

$$\frac{\partial C}{\partial t} + \nabla \cdot \left[ \left( \sum_\alpha \mathbf{e}_\alpha g_\alpha^{eq} \right) + \left(1 - \frac{1}{2\tau_g}\right) \left( \sum_\alpha \mathbf{e}_\alpha g_\alpha^{neq} \right) \right] = 0, \quad (31)$$

where $f_\alpha^{neq-m}(\mathbf{x},t) = -\tau_f \left[ f_\alpha^{eq}(\mathbf{x},t) - f_\alpha^{eq-m}(\mathbf{x} - \mathbf{e}_\alpha \Delta t, t - \Delta t) \right]$ is the non-equilibrium term for the flow field and $g_\alpha^{neq}(\mathbf{x},t) = -\tau_g \left[ g_\alpha^{eq}(\mathbf{x},t) - g_\alpha^{eq}(\mathbf{x} - \mathbf{e}_\alpha \delta_t, t - \delta_t) \right]$ is the non-equilibrium term for the phase field. Define $f_\alpha^*$ and $g_\alpha^*$ as:

$$f_\alpha^* = f_\alpha^{eq} - \left(1 - \frac{1}{2\tau_f}\right) f_\alpha^{neq\_m}, \quad (32)$$

$$g_\alpha^* = \left[ g_\alpha^{eq} + \left(1 - \frac{1}{2\tau_g}\right) g_\alpha^{neq} \right]. \quad (33)$$

As a result, the relations between the macroscopic fluxes and the distribution functions

can be derived as:

$$\begin{cases} \rho \mathbf{u} c_s^2 = \sum_\alpha \mathbf{e}_\alpha f_\alpha^{eq} \\ \rho \mathbf{u}\mathbf{u} + p\mathbf{I} + \mu\left(\nabla \mathbf{u} + (\nabla \mathbf{u})^T\right) = \dfrac{1}{c_s^2}\left[\sum_\alpha \mathbf{e}_\alpha \mathbf{e}_\alpha f_\alpha^*\right] \end{cases}, \quad (34)$$

$$C\mathbf{u} - M_c \nabla \mu_c = \sum_\alpha \mathbf{e}_\alpha g_\alpha^*. \quad (35)$$

Finally, the macroscopic equations for the flow and phase fields can be expressed as:

$$\begin{cases} \dfrac{\partial p}{\partial t} + \nabla \cdot \left(\sum_\alpha \mathbf{e}_\alpha f_\alpha^{eq}\right) - \mathbf{u} \cdot \nabla \rho c_s^2 - \rho c_s^2 \left(1 - \dfrac{\rho_s}{\rho_l}\right)\dfrac{\partial F_s}{\partial t} = 0 \\ \dfrac{\partial \rho c_s^2 \mathbf{u}}{\partial t} + \nabla \cdot \left(\sum_\alpha \mathbf{e}_\alpha \mathbf{e}_\alpha f_\alpha^*\right) - c_s^2 \left(\mathbf{F}_{surf} + \mathbf{F}_g + \mathbf{F}_{mg} + \mathbf{F}_{IBM}\right) = 0 \end{cases}, \quad (36)$$

$$\dfrac{\partial C}{\partial t} + \nabla \cdot \sum_\alpha \mathbf{e}_\alpha g_\alpha^* - C\left(1 - \dfrac{\rho_s}{\rho_l}\right)\dfrac{\partial F_s}{\partial t} = 0. \quad (37)$$

### 3.4 Finite volume discretization and time evolution

The macroscopic equations for the magnetic, temperature, flow, and phase fields are discretized using a cell-centered finite volume method (FVM). Eqs. (17), (25), (36), and (37) are expressed as the following unified form:

$$\dfrac{\partial \mathbf{W}}{\partial t} + \nabla \cdot \mathbf{F} = \mathbf{F}_E, \quad (38)$$

where

$$\mathbf{W} = \begin{pmatrix} \phi_m \\ H_e \\ p \\ \rho c_s^2 u \\ \rho c_s^2 v \\ C \end{pmatrix}, \quad (39)$$

$$\mathbf{F} = \begin{pmatrix} \sum_\alpha \mathbf{e}_\alpha m_\alpha^* \\ \sum_\alpha \mathbf{e}_\alpha h_\alpha^* \\ \sum_\alpha \mathbf{e}_\alpha f_\alpha^{eq} \\ \sum_\alpha \mathbf{e}_\alpha e_{\alpha x} f_\alpha^* \\ \sum_\alpha \mathbf{e}_\alpha e_{\alpha y} f_\alpha^* \\ \sum_\alpha \mathbf{e}_\alpha g_\alpha^* \end{pmatrix}, \tag{40}$$

$$\mathbf{F}_E = \begin{pmatrix} 0 \\ C_p T \nabla \cdot \mathbf{u} \\ \mathbf{u} \cdot \nabla \rho c_s^2 + \rho c_s^2 \left(1 - \dfrac{\rho_s}{\rho_l}\right) \dfrac{\partial F_s}{\partial t} \\ c_s^2 \left(F_{surf,x} + F_{mg,x} + F_{IBM,x}\right) \\ c_s^2 \left(F_{surf,y} + F_{mg,y} + F_{g,y} + F_{IBM,y}\right) \\ C \left(1 - \dfrac{\rho_s}{\rho_l}\right) \dfrac{\partial F_s}{\partial t} \end{pmatrix}. \tag{41}$$

From Eq. (40), it is obvious that the macroscopic fluxes can be calculated from the moments of distribution functions at the cell interface, where the standard LBE is satisfied. By integrating Eq. (38) over a control volume $i$, the following semi-discretized form of the macroscopic governing equations can be obtained:

$$\frac{d\mathbf{W}_i}{dt} = -\frac{1}{V_i} \sum_{j \in N(i)} \left(n_x \mathbf{F}_x + n_y \mathbf{F}_y\right)_{ij} S_{ij} + \mathbf{F}_{E,i}, \tag{42}$$

where $V_i$, $N(i)$, and $S_{ij}$ are the volume of control cell $i$, the set of neighboring cells of cell $i$, and the area of the interface shared between cell $i$ and cell $j$, respectively. $\mathbf{n}_{ij}=(n_x, n_y)$ represents the unit outer normal vector at the cell surface. $\mathbf{F}_x$ and $\mathbf{F}_y$ can be expressed as:

$$\mathbf{F}_x = \begin{pmatrix} \sum_\alpha e_{\alpha x} m_\alpha^* \\ \sum_\alpha e_{\alpha x} h_\alpha^* \\ \sum_\alpha e_{\alpha x} f_\alpha^{eq} \\ \sum_\alpha e_{\alpha x} e_{\alpha x} f_\alpha^* \\ \sum_\alpha e_{\alpha x} e_{\alpha y} f_\alpha^* \\ \sum_\alpha e_{\alpha x} g_\alpha^* \end{pmatrix}, \quad \mathbf{F}_y = \begin{pmatrix} \sum_\alpha e_{\alpha y} m_\alpha^* \\ \sum_\alpha e_{\alpha y} h_\alpha^* \\ \sum_\alpha e_{\alpha y} f_\alpha^{eq} \\ \sum_\alpha e_{\alpha y} e_{\alpha x} f_\alpha^* \\ \sum_\alpha e_{\alpha y} e_{\alpha y} f_\alpha^* \\ \sum_\alpha e_{\alpha y} g_\alpha^* \end{pmatrix}. \qquad (43)$$

Consider the interface between two adjacent control volumes shown in Fig. 2, the distribution functions in Eq. (43) are evaluated at the middle point of the interface. Furthermore, these distribution functions depend on the equilibrium states at the cell interface and its surrounding points.

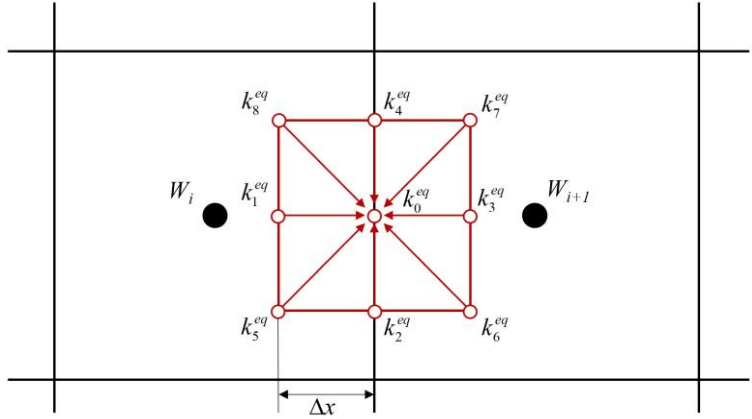

Fig. 2. Local reconstruction of LBE solution at a cell interface.

In the cell-centered FVM, all macroscopic variables are defined at the cell centers. To obtain their values at the surrounding points of the cell interface, the following interpolation method is employed:

$$\psi(\mathbf{x}-\mathbf{e}_\alpha \delta_t, t-\delta_t) = \begin{cases} \psi(\mathbf{x}_i) + (\mathbf{x}-\mathbf{e}_\alpha \delta_t - \mathbf{x}_i)\cdot \nabla \psi(\mathbf{x}_i), & (\mathbf{x}-\mathbf{e}_\alpha \Delta t) \in W_i \\ \psi(\mathbf{x}_{i+1}) + (\mathbf{x}-\mathbf{e}_\alpha \delta_t - \mathbf{x}_{i+1})\cdot \nabla \psi(\mathbf{x}_{i+1}), & (\mathbf{x}-\mathbf{e}_\alpha \Delta t) \in W_{i+1} \end{cases} \qquad (44)$$

where $\mathbf{x}_i$, $\mathbf{x}_{i+1}$, and $\mathbf{x}$ represent the physical positions of two adjacent cell centers and their shared interface, respectively. $\psi$ denotes the physical variables such as $\phi_m$, $H$, $p$, $\mathbf{u}$, and $C$. Note that for the variables at the cell interface at the current time level, their values are taken as the average of the interpolated results from the left and right cells. After interpolation, the equilibrium states at the surrounding points of the cell interface $m_\alpha^{eq}(\mathbf{x}-\mathbf{e}_\alpha\delta_t, t-\delta_t)$, $h_\alpha^{eq}(\mathbf{x}-\mathbf{e}_\alpha\delta_t, t-\delta_t)$, $f_\alpha^{eq}(\mathbf{x}-\mathbf{e}_\alpha\delta_t, t-\delta_t)$ and $g_\alpha^{eq}(\mathbf{x}-\mathbf{e}_\alpha\delta_t, t-\delta_t)$ can be obtained through Eqs. (13), (20), (28), and (29), and the newly defined distribution function $f_\alpha^{eq-m}(\mathbf{x}-\mathbf{e}_\alpha\delta_t, t-\delta_t)$ can be calculated. According to the compatibility conditions, the macroscopic variables at the cell interface at the new time level can be computed from the surrounding equilibrium states:

$$\begin{cases} \phi_m(\mathbf{x},t) = \sum_\alpha m_\alpha^{eq}(\mathbf{x},t) = \sum_\alpha m_\alpha^{eq}(\mathbf{x}-\mathbf{e}_\alpha\delta_t, t-\delta_t) \\ H_e(\mathbf{x},t) = \sum_\alpha h_\alpha^{eq}(\mathbf{x},t) = \sum_\alpha h_\alpha^{eq}(\mathbf{x}-\mathbf{e}_\alpha\delta_t, t-\delta_t) \\ p(\mathbf{x},t) = \sum_\alpha f_\alpha^{eq}(\mathbf{x},t) = \sum_\alpha f_\alpha^{eq}(\mathbf{x}-\mathbf{e}_\alpha\delta_t, t-\delta_t) \\ \rho(\mathbf{x},t)\mathbf{u}(\mathbf{x},t)c_s^2 = \sum_\alpha \mathbf{e}_\alpha f_\alpha^{eq}(\mathbf{x},t) = \sum_\alpha \mathbf{e}_\alpha f_\alpha^{eq}(\mathbf{x}-\mathbf{e}_\alpha\delta_t, t-\delta_t) \\ C(\mathbf{x},t) = \sum_\alpha g_\alpha^{eq}(\mathbf{x},t) = \sum_\alpha g_\alpha^{eq}(\mathbf{x}-\mathbf{e}_\alpha\delta_t, t-\delta_t) \end{cases} \quad (45)$$

The source terms in Eq. (41) can be directly calculated at the cell centers. Once the macroscopic fluxes and the source terms are determined, Eq. (42) can be simplified as:

$$\frac{\partial \mathbf{W}_i}{\partial t} = \mathbf{R}(\mathbf{W}) \quad (46)$$

where $\mathbf{R}(\mathbf{W})$ encompasses the fluxes $\nabla \cdot \mathbf{F}$ and the source term $\mathbf{F}_E$. The above equation can be solved using the third-order Runge-Kutta method [55]. The whole computational process of the proposed solver is summarized in Table 1.

Table 1. Computational process of the proposed solver.

| **Algorithm:** Computational procedure of the proposed solver |
| --- |
| **Initialization** |
| Initialize the physical variables $\phi_m$, $H_e$, $p$, $\mathbf{u}$, $\rho$, $C$, and $\mu_c$; |
| **End Initialization** |
| **Main Iteration Loop** |

(a) Specify the streaming time step $\delta_t$. Ensure that the surrounding points $(\mathbf{x} - \mathbf{e}_\alpha \delta_t)$ are within an appropriate range of the target cells $W_i$ and $W_{i+1}$. Then, calculate the single-relaxation-time parameters $\tau_m$, $\tau_h$, $\tau_f$, $\tau_g$;

(b) Calculate the macroscopic quantities $\phi_m$ at $(\mathbf{x} - \mathbf{e}_\alpha \delta_t)$ by Eq. (44), and compute the equilibrium states $m_\alpha^{eq}(\mathbf{x} - \mathbf{e}_\alpha \delta_t, t - \delta_t)$ by Eq. (13). Determine the macroscopic variables at the midpoint of the interface between two adjacent cells by Eq. (45), and then evaluate the equilibrium states $m_\alpha^{eq}(\mathbf{x}, t)$ by Eq. (13);

(c) Evaluate the non-equilibrium term $m_\alpha^{neq}$, and obtain $m_\alpha^*$ by Eq. (16). Solve the semi-discretized macroscopic magnetic potential equation (17) using the third-order Runge-Kutta method, and calculate $F_{mgx}$ and $F_{mgy}$;

**while** *Magnetic field is convergent* **do**

    (a) Calculate the macroscopic quantities $H_e$, $p$, $\mathbf{u}$, $C$, $\rho$, and $\mu_c$ at $(\mathbf{x} - \mathbf{e}_\alpha \delta_t)$ by Eq. (44), and obtain the equilibrium states $h_\alpha^{eq}(\mathbf{x} - \mathbf{e}_\alpha \delta_t, t - \delta_t)$ $f_\alpha^{eq}(\mathbf{x} - \mathbf{e}_\alpha \delta_t, t - \delta_t)$, $f_\alpha^{eq-m}(\mathbf{x} - \mathbf{e}_\alpha \delta_t, t - \delta_t)$, and $g_\alpha^{eq}(\mathbf{x} - \mathbf{e}_\alpha \delta_t, t - \delta_t)$ by Eqs. (20), (28), and (29);

    (b) Compute the macroscopic quantities at the midpoint of the interface between two adjacent cells by Eq. (45), and evaluate the equilibrium states $h_\alpha^{eq}(\mathbf{x}, t)$, $f_\alpha^{eq}(\mathbf{x}, t)$ and $g_\alpha^{eq}(\mathbf{x}, t)$ by Eqs. (20), (28), and (29);

    (c) Determine the non-equilibrium terms $h_\alpha^{neq}$, $f_\alpha^{neq-m}$, and $g_\alpha^{neq}$, and obtain $h_\alpha^*$, $f_\alpha^*$, and $g_\alpha^*$ by Eqs. (22), (32), and (33). Calculate the fluxes $\mathbf{F}$ at the cell interface and the source terms $\mathbf{F}_E$ at the cell center. Solve the semi-discretized total enthalpy, NS, and CH equations using the third-order Runge-Kutta method.

    (d) Update the temperature $T$ and liquid fraction $F_l$ via Eq. (8).

**end**

**if** *Output required* **then**

    Write global parameter data to a file.

**end**

**End Main Iteration Loop**

## 4. Model validation

In this section, several benchmark cases are conducted to validate the accuracy and reliability of the proposed solver, including conductive freezing, static droplet freezing, and ferrofluid droplet deformation. The simulation results obtained using the proposed lattice Boltzmann flux solver are compared with available theoretical and experimental data from previous studies.

### 4.1 Conductive freezing

To verify the accuracy of the temperature field during solidification, the classical conductive freezing problem is considered, as illustrated in Fig. 3. The computational domain initially contains liquid at temperature $T_0$. The left wall is maintained at a constant temperature $T_b$ ($T_b < T_m$), while the right wall remains at $T_0$ ($T_0 > T_m$), where $T_m$ is the melting temperature. The velocity field is set to zero, with Dirichlet boundary conditions in the x-direction and periodic boundary conditions in the y-direction. The analytical temperature field is given by [32]:

$$T(X,t) = \begin{cases} T_b - \dfrac{(T_b - T_m)\,\mathrm{erf}\left(\dfrac{x}{2\sqrt{\alpha_s t}}\right)}{\mathrm{erf}(k)}, & 0 < X < X_i(t) \\[2mm] T_0 + \dfrac{(T_m - T_b)\,\mathrm{erf}\left(\dfrac{x}{2\sqrt{\alpha_l t}}\right)}{\mathrm{erf}\left(k\sqrt{\dfrac{\alpha_l}{\alpha_s}}\right)}, & X > X_i(t) \end{cases}, \qquad (47)$$

where $X_i(t) = 2k\sqrt{\alpha_s t}$ is the position of the solid-liquid interface, and $k$ can be obtained by the following equation:

$$\frac{C_{p,s}(T_m - T_b)}{L\exp(k^2)\,\mathrm{erf}(k)} - \frac{C_{p,l}(T_0 - T_m)\sqrt{\dfrac{\alpha_l}{\alpha_s}}}{L\exp\left(\dfrac{k^2 \alpha_s}{\alpha_l}\right)\mathrm{erf}\left(k\sqrt{\dfrac{\alpha_s}{\alpha_l}}\right)} = k\sqrt{\pi} \qquad (48)$$

The physical parameters are set as $\rho_l = \rho_s = 1$, $C_{p,s}=C_{p,l}=1$, $\alpha_s = \alpha_l = 0.4$, $L=250$, $T_b=-1$, $T_0=1$, and $T_m=0$. Fig. 4 compares the temperature profiles from theoretical predictions and numerical simulations at t=5, t=10, and t=15. The excellent agreement demonstrates that the proposed solver can accurately reproduce heat conduction and solidification in a pure material.

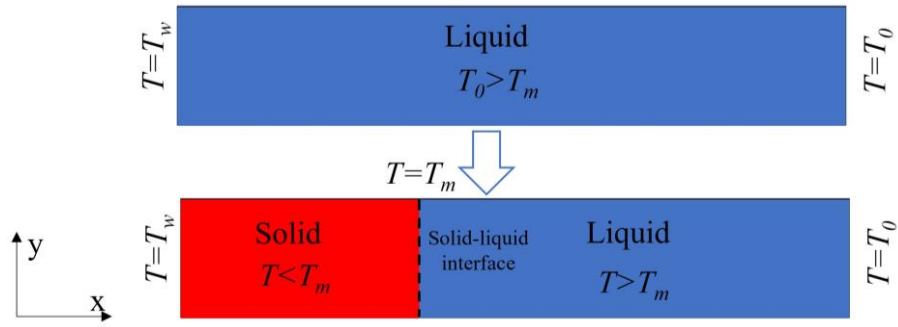

Fig. 3. Schematic of the conductive freezing problem.

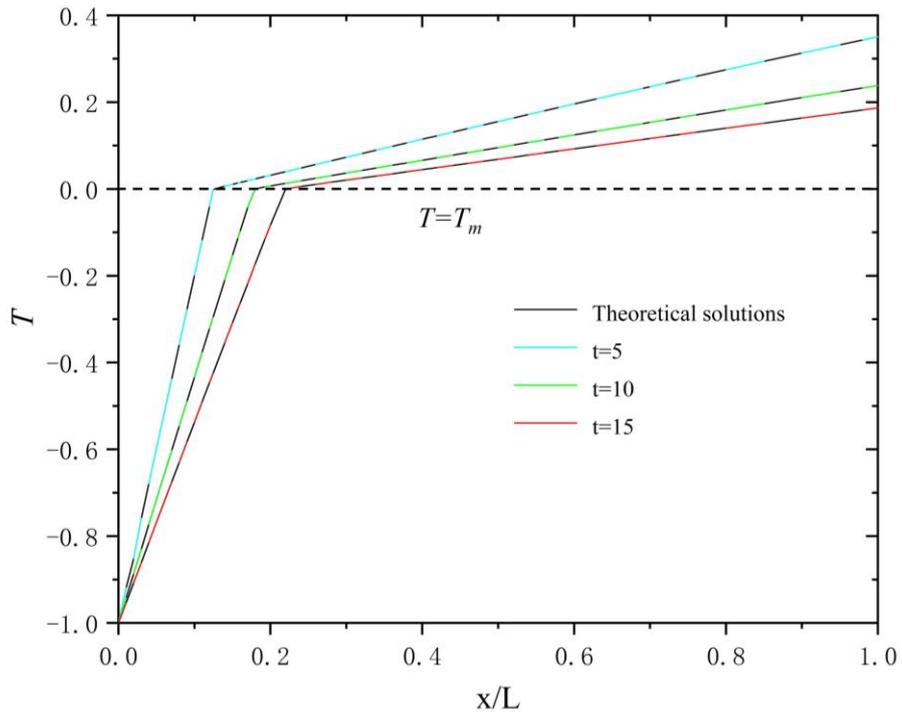

Fig. 4. Comparison between theoretical and simulated temperature distributions at different times.

## 4.2 A static droplet freezing

To further assess the solver's ability to capture phase change dynamics, we simulate the freezing process of a water droplet on a cold surface at two different contact angles $\theta = 64°$ and $\theta = 98°$. The dimensionless parameters employed are the Stefan number ( $Ste = C_{p,l}(T_m - T_w)/L = 0.19$ ), Prandtl number ( $Pr = \mu_l C_{p,l}/\lambda_l = 13.47$ ), and Bond number ( $Bo = gR_0(\rho_l - \rho_g)/\sigma = 0.22$ ). The density ratio between the solid, liquid, and gas phases is 0.9, 1, and 0.05, and the droplet volume is 4.5 $\mu L$. The initial droplet temperature is $T_m$=0°C, and the bottom wall temperature is fixed at $T_w$=–15°C. Adiabatic boundary conditions are applied to the left, right, and top walls for the temperature field, while a Dirichlet condition ($T=T_w$) is imposed at the bottom. The flow field satisfies no-slip conditions at all walls, and the phase field enforces no-flux boundaries. The computational domain is discretized on a 400 × 200 grid.

Fig. 5 illustrates the schematic of the droplet freezing process, with the solid length ($R_l$) and height ($H_l$) defined as shown. Fig. 6 presents a comparison between the simulated and experimental freezing interfaces [33]. The black line denotes the solid–liquid boundary. At the onset of freezing, the interface advances upward. In the experiments, the solidification front appears relatively flat due to optical limitations caused by opaque ice. However, the simulations reveal a slightly concave interface, consistent with observations by Marin et al. [57]. Notably, the simulations capture the outward expansion of the interface and the formation of a conical tip at the top of the droplet, driven by the lower density of ice and surface tension. These results confirm that the proposed solver can accurately resolve the transient solidification behavior of water droplets.

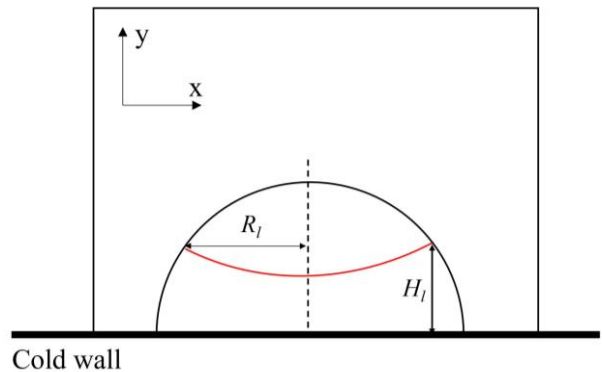

Fig. 5. Diagram of a droplet freezing on the cold wall and definitions of $R_l$ and $H_l$.

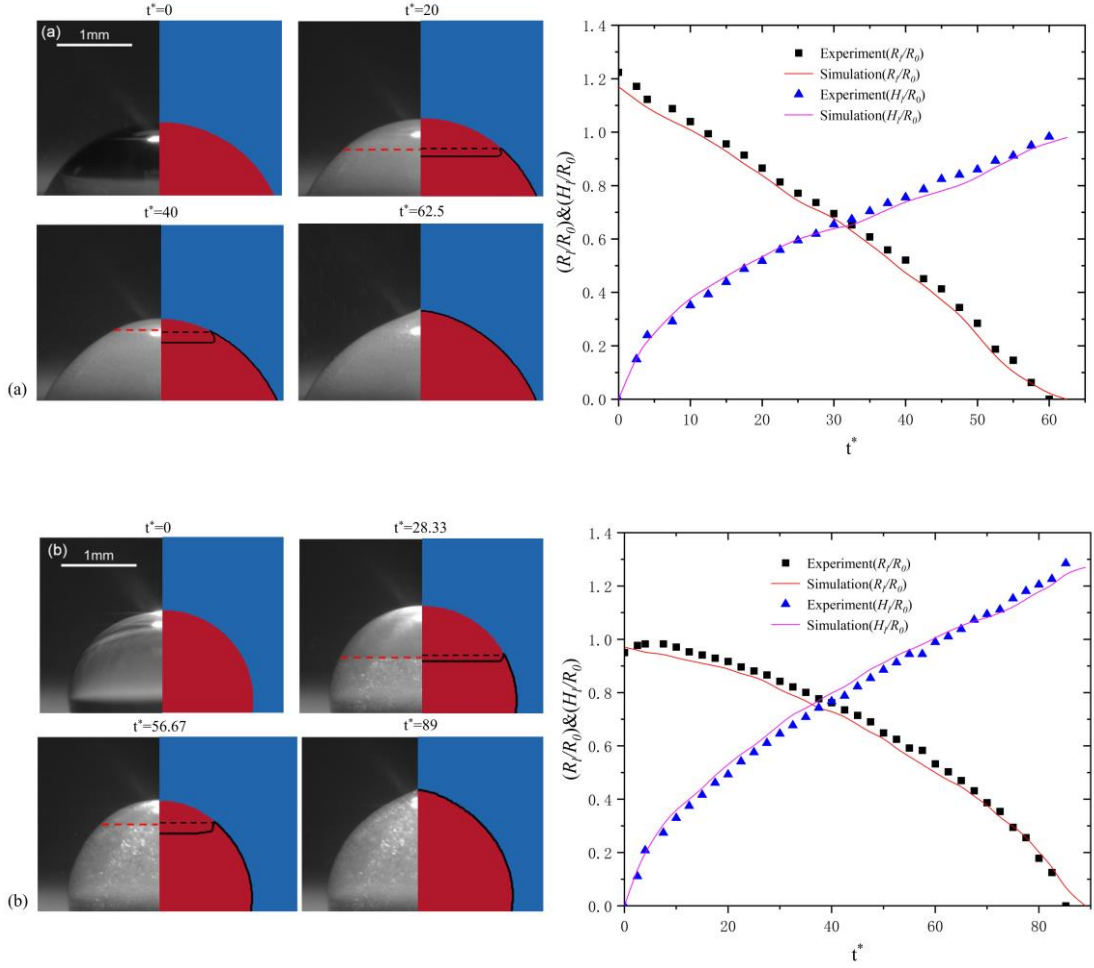

Fig. 6. The freezing process of the water droplet at (a), and the comparison of the experimental result and the simulation result of $H_l$ and $R_l$.

## 4.3 Deformation of a ferrofluid droplet

Finally, to validate the magnetic field module, we examine the deformation of a water-based ferrofluid droplet suspended in an organic oil under a uniform magnetic field. The physical properties of the ferrofluid and carrier oil are listed in Table 2. The droplet has an initial diameter of $R_0 = 1$ mm and a nanoparticle volume fraction of 10% [47]. No-slip boundary conditions are applied to all walls. A uniform magnetic field is imposed along the top and bottom boundaries, while magnetic insulation conditions are applied to the left and right boundaries. The computational domain is discretized on a 200 × 200 grid. Fig. 7 compares the equilibrium droplet shapes under five different magnetic field intensities. The deformation ratio (b/a) from our simulations agrees closely with the experimental data of Flament et al. [58]. This strong agreement demonstrates that the proposed solver can accurately capture the magnetically induced deformation of ferrofluid droplets.

Table 2. Physical parameters of the ferrofluid droplet and surrounding fluid.

| Property | Ferrofluid | Organic oil |
|---|---|---|
| Viscosity, $\mu(Pa\ s)$ | 0.016 | 0.0008 |
| Surface tension, $\sigma(N/m)$ | 3.07 | - |
| Density, $\rho(kg/m^3)$ | 1580 | 800 |
| Saturation magnetization, $M_s(kA/m)$ | 40 | - |
| Magnetic permeability, $\mu_m$ | 2.2 | - |

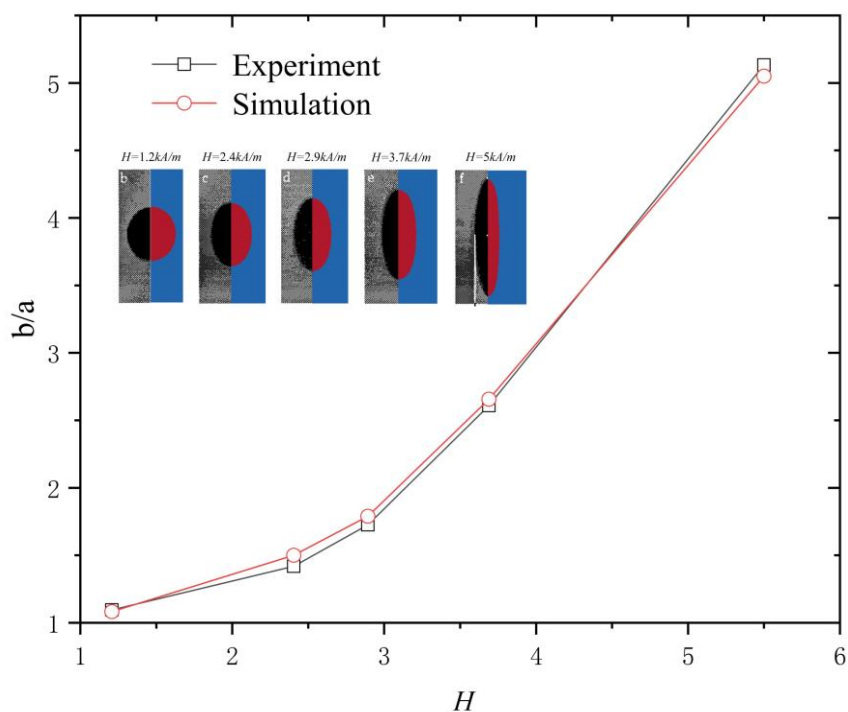

Fig. 7. Comparison between experimental and simulated deformation ratios b/a under different magnetic field strengths.

## 5. Results and discussion

This section investigates the effect of a uniform magnetic field on the freezing dynamics of a water-based ferrofluid droplet. The schematic configuration of the study is shown in Fig. 8. The uniform magnetic field is applied in two orientations: (a) vertical and (b) horizontal. The contact angle is denoted as $\theta = 90°$. Thermodynamically, droplet freezing on a cooled substrate involves four distinct stages: supercooling, crystal nucleation, equilibrium freezing, and solid supercooling [59]. This process presents a multiscale challenge because the time-scale difference between nucleation and solidification exceeds three orders of magnitude [42]. Such disparity makes full-scale simulations of ice crystal growth computationally prohibitive, even with high-performance computing resources. Consequently, prior studies [28], [56] typically neglect the nucleation stage and treat the pre-nucleation period as an initial condition. In this work, the nucleation stage is similarly omitted.

The boundary conditions are as follows: the temperature field adopts adiabatic conditions at the left, right, and top walls and a Dirichlet condition ($T_w$) at the bottom. The flow field employs no-slip conditions on all walls, and the phase field uses no-flux conditions along all domain boundaries. The physical parameters of the ferrofluid droplet are summarized in Table 3. Its effective thermal conductivity and diffusivity are determined by the Maxwell model [60]:

$$\frac{\Omega_e}{\Omega_l} = 1 + \frac{3(n-1)\varphi}{(n+2)-(n-1)\varphi}, \tag{49}$$

where $\varphi$ is the volume fraction of $Fe_3O_4$ particles, and $\Omega_e$ and $\Omega_l$ are the effective property and property of the continuous phase, respectively. $n = \Omega_f / \Omega_l$, where $\Omega_f$ is the property of $Fe_3O_4$ particles. The latent heat of the ferrofluid is expressed as $L_f = (1-\varphi)L_w$, where $L_w$ is the latent heat of pure water. Several dimensionless numbers are adopted to characterize the freezing process of the ferrofluid droplet: the Stefan number (*Ste*), Fourier number (*Fo*), Peclet number (*Pe*), and magnetic Bond number (*Bo_m*), which are defined as:

$$Ste = \frac{C_{p,l}(T_m - T_w)}{L}, Fo = \frac{\lambda_l t}{\rho_l C_{p,l} R_0^2}, Pe = \frac{Du^2}{\alpha}, Bo_m = \frac{\mu_0 H^2 D}{\sigma}, \tag{50}$$

where *Ste* represents the ratio of sensible to latent heat, *Fo* is the dimensionless time, *Pe* measures convective strength caused by the uniform magnetic field, and $Bo_m$ characterizes the strength of the applied magnetic field.

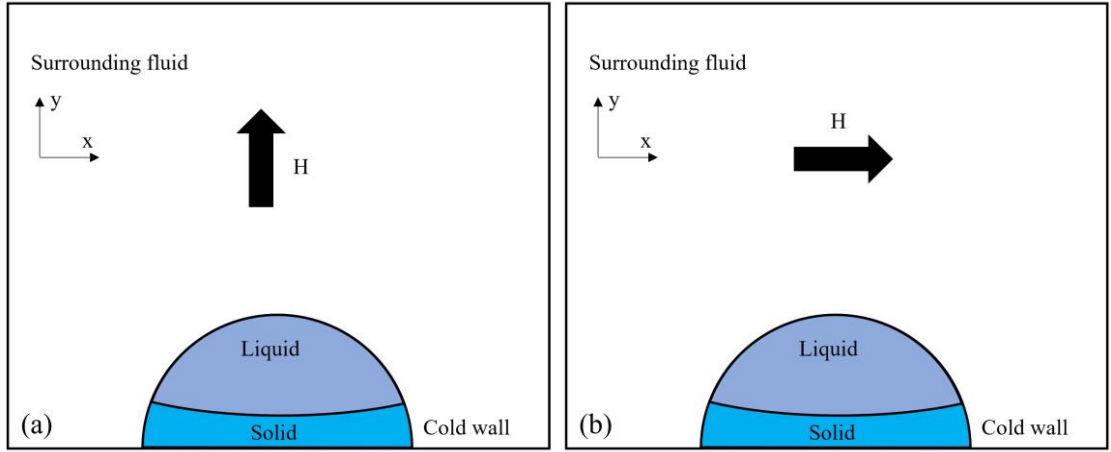

Fig. 8. Schematic of a ferrofluid droplet freezing on a cooled substrate under (a) vertical and (b) horizontal uniform magnetic fields.

Table 3. Physical parameters of the ferrofluid droplet.

| Property | Ferrofluid | |
|---|---|---|
| | Liquid | Solid |
| Viscosity, $\mu(Pa\ s)$ | 0.016 | - |
| Surface tension, $\sigma(N/m)$ | 0.00307 | - |
| Density, $\rho(kg/m^3)$ | 1580 | 1450 |
| Specific heat capacity, $C_p(J/(kg \cdot K))$ | 3000 | 1670 |
| Thermal conductivity, $\lambda(W/(m \cdot K))$ | 0.72 | 2.32 |
| Latent heat, $L(J/kg)$ | 300009 | - |
| Solidification temperature, $T_m(°C)$ | 0 | - |

## 5.1 Effect of uniform magnetic field on freezing morphology

The morphological evolution of the ferrofluid droplet during freezing under uniform magnetic fields was first examined. Numerical simulations were performed for droplets subjected to both vertical and horizontal magnetic fields of varying intensities. As shown in Fig. 9, the solidification front progresses from the bottom to the top of the droplet. Due to the higher thermal diffusivity of the surrounding environment, the droplet edges freeze earlier than the center, producing a relatively flat freezing front. This behavior corresponds to the secondary freezing behavior reported by Mohammadipour et al. [28]. Moreover, under a vertical magnetic field, the droplet elongates along the field direction, increasing the overall freezing time. In contrast, a horizontal magnetic field flattens the droplet, resulting in a shorter freezing duration. To further elucidate the role of magnetic field orientation, Fig. 10 presents the distribution of magnetic forces acting on the droplet for both field orientations. The magnetic force acts on the droplet interface and vanishes within the bulk region. In a vertical field, the poles experience strong upward magnetic forces, whereas the equatorial region experiences weaker forces, leading to elongation. Conversely, in a horizontal field, the equatorial poles experience maximum magnetic forces, flattening the droplet along the field direction.

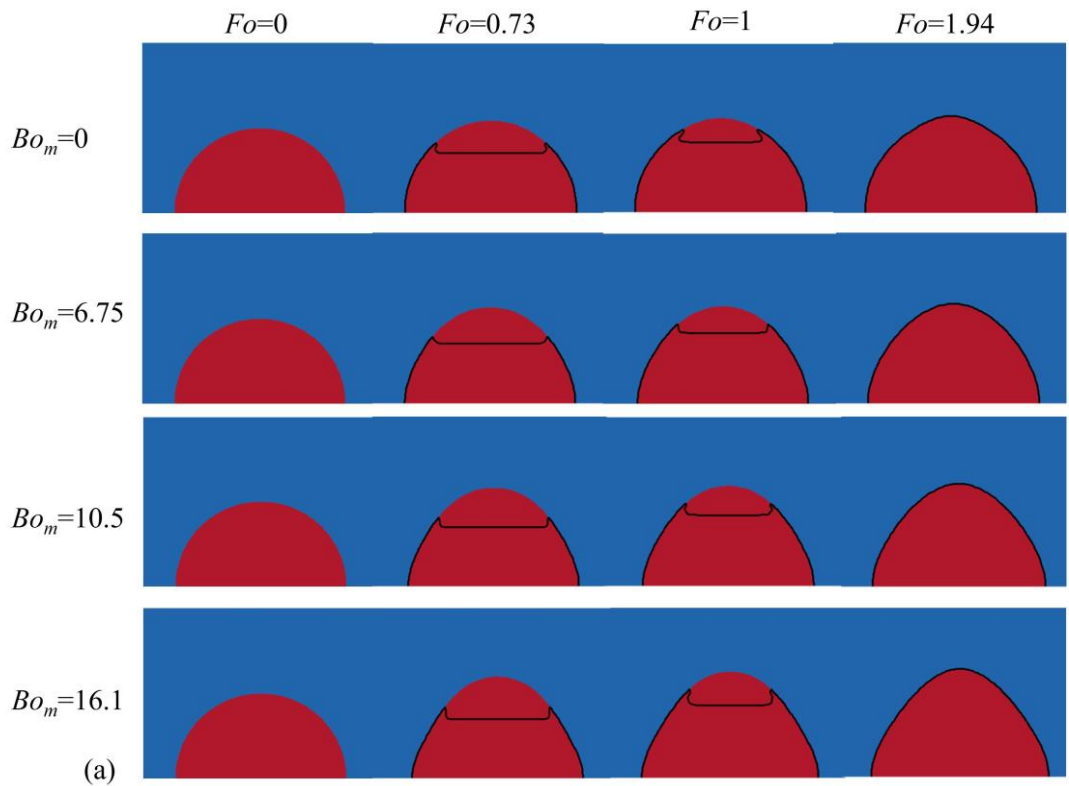

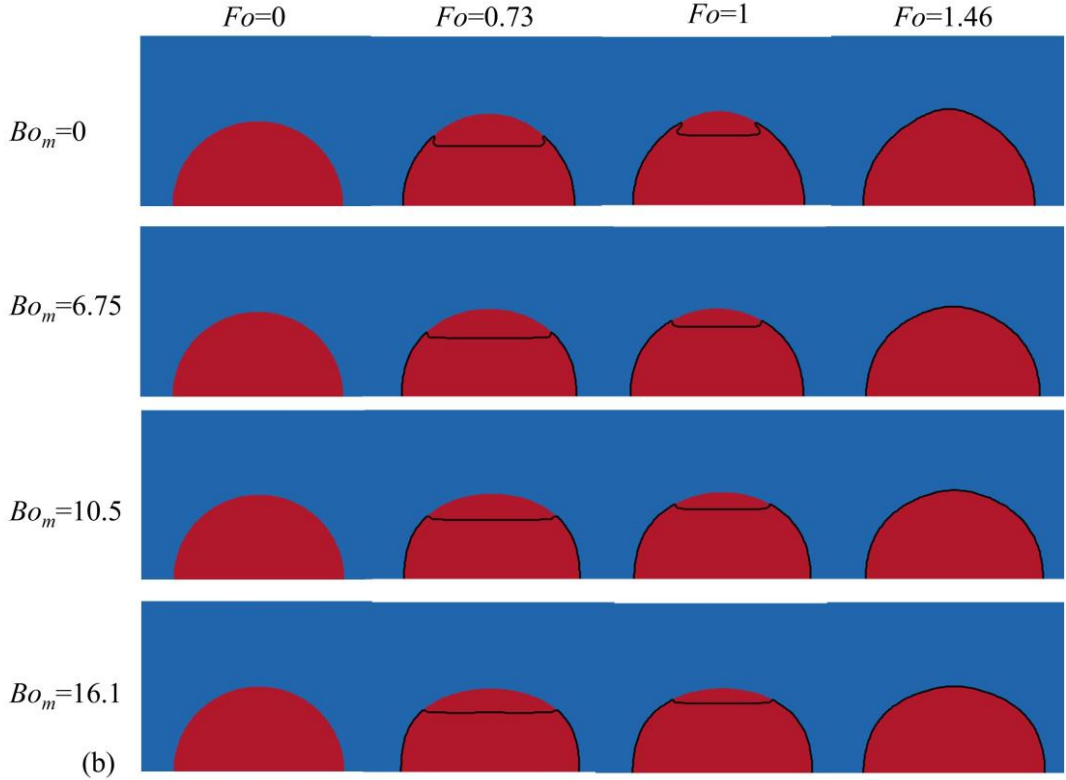

Fig. 9. Evolution of the freezing front (black line) under (a) vertical and (b) horizontal magnetic fields for different $Bo_m$ with $Ste$=0.3.

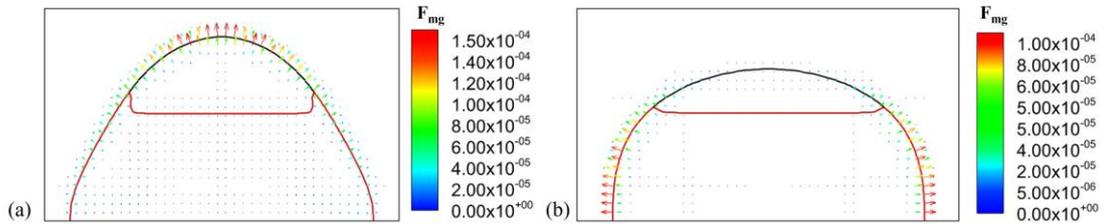

Fig. 10. Magnetic force distributions at the droplet interface under (a) vertical and (b) horizontal fields for $Bo_m$=16.1 and $Fo$=0.73.

Fig. 11 shows the time evolution of the normalized droplet height $H_m/R_0$ under vertical and horizontal magnetic fields. It should be noted that the abbreviations VMF and HMF in all figures refer to the vertical uniform magnetic field and horizontal uniform magnetic field, respectively. In the absence of a magnetic field, the normalized height $H_m/R_0$ gradually increases during solidification until complete freezing occurs. This increase results from the lower density of the solid phase relative to the liquid phase, which leads to volumetric expansion upon freezing. When a uniform magnetic field is applied, the droplet exhibits distinct morphological responses depending on the field orientation. The magnetic force associated with the vertical field elongates the droplet along the field direction, while the horizontal field compresses and flattens it in the same direction. As the magnetic field intensity increases, the normalized height $H_m/R_0$ under the vertical field rises, reaching a maximum enhancement of approximately 10.8%. In contrast, under the horizontal field, $H_m/R_0$ decreases with increasing field strength, with a maximum reduction of about 10.1%. It is also noteworthy that, at the early stage of freezing ($Fo < 0.273$), the magnetic field exerts a more pronounced influence, resulting in observable fluctuations in $H_m/R_0$. As solidification progresses, the magnetic effect gradually diminishes, indicating that magnetic deformation primarily occurs during the initial stage of the freezing process.

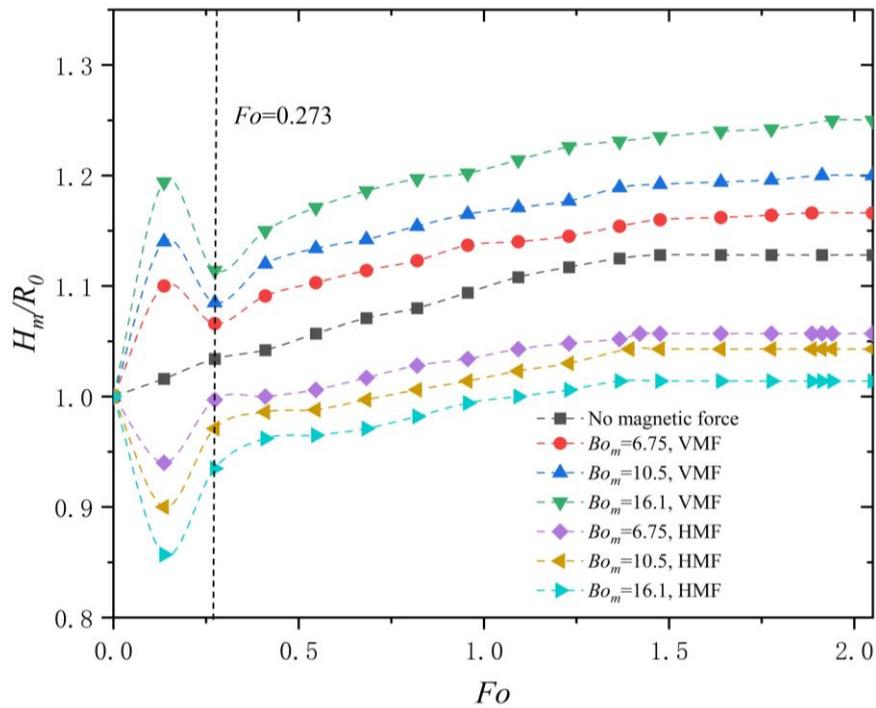

Fig. 11. Evolution of normalized height $H_m/R_0$ under different magnetic field orientations.

**5.2 Effect of uniform magnetic field on heat transfer**

This section investigates the influence of a uniform magnetic field on heat transfer during the droplet freezing process. Fig. 12 shows the evolution of the Péclet number ($Pe$) with dimensionless time ($Fo$) under various magnetic field strengths. The results indicate that the application of a uniform magnetic field markedly enhances internal convection within the droplet. However, the values of *Pe* remain much smaller than 1 in all cases, confirming that convective heat transfer is negligible compared with thermal conduction. Fig. 13 further illustrates the interfacial flow field during freezing. It can be seen that when a uniform magnetic field is applied, the interfacial velocity increases significantly compared with the case without a magnetic field. This enhanced interfacial motion induces internal circulation within the droplet, thereby slightly augmenting convective heat transport. Additionally, Fig. 14 presents the temperature distributions within the droplet and along its centerline. The results show that the temperature profiles along the droplet centerline remain nearly unchanged after applying the magnetic field, demonstrating that conductive heat transfer dominates and that convective effects are minimal at *Ste* = 0.3. According to Starostin et al. [61], the spatial orientation of the freezing cone is primarily determined by the temperature gradient direction at the droplet–substrate interface. The present findings further confirm that the droplet tips remain symmetrical under a uniform magnetic field, implying that the field does not alter the heat flow direction during solidification.

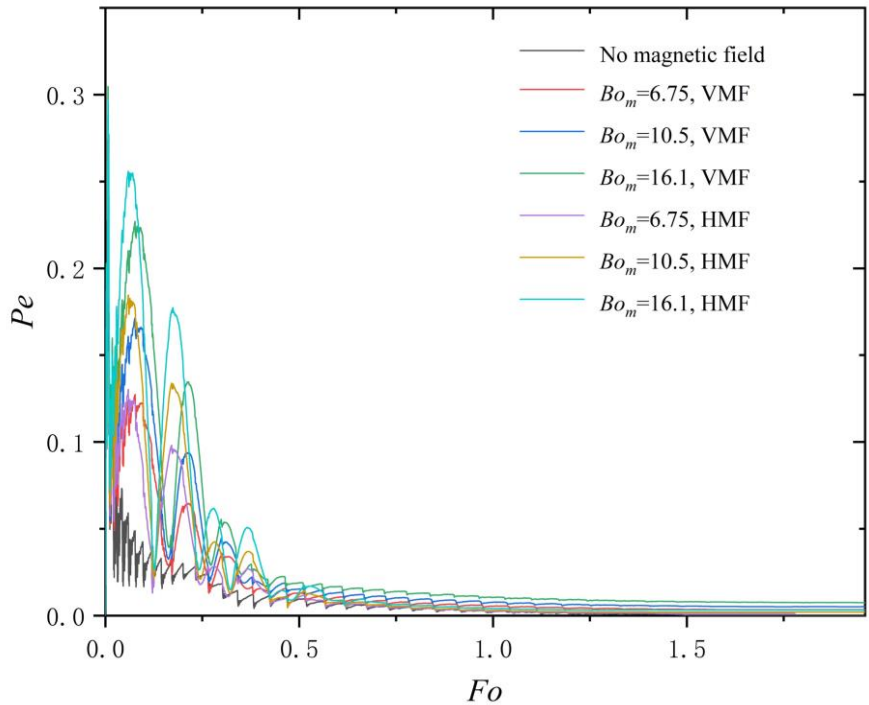

Fig. 12. Evolution of *Pe* with *Fo* under different magnetic field strengths.

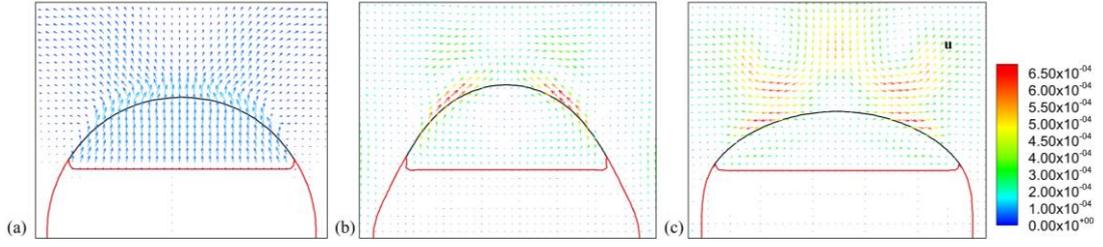

Fig. 13. Distribution of interfacial flow fields under different $Bo_m$ numbers. (a) without magnetic force, (b) $Bo_m$=16.1, vertical uniform magnetic field, (c) $Bo_m$=16.1, horizontal uniform magnetic field at $Fo$=0.68.

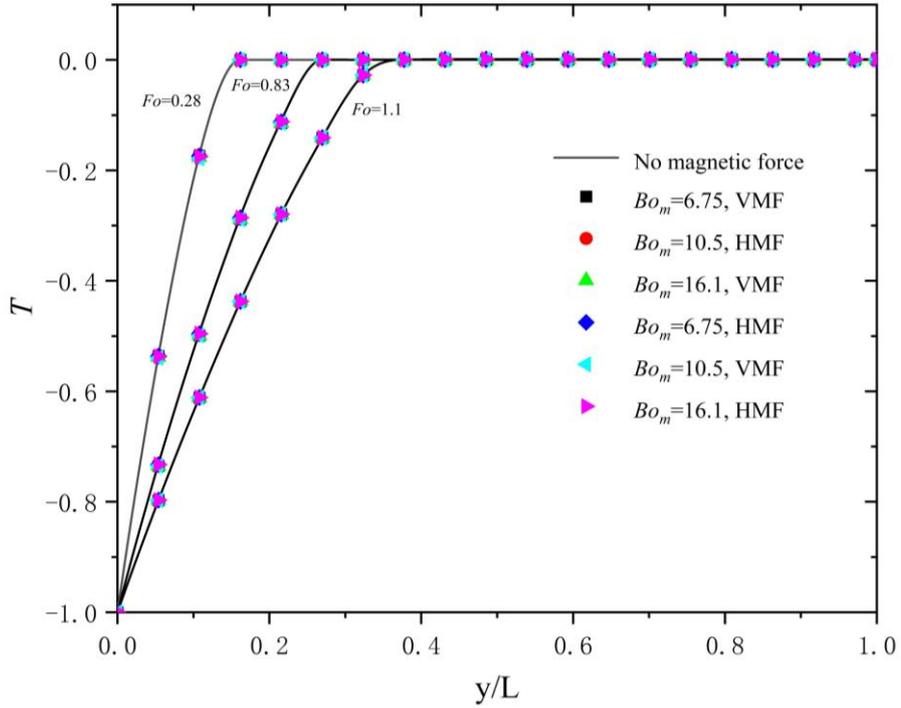

Fig. 14. Temperature distribution within the droplet and along its centerline under (a) vertical and (b) horizontal magnetic fields at different $Fo$ numbers.

Fig.15 shows the local heat flux on the cold surface at different $Fo$s under vertical and horizontal uniform magnetic fields. The local heat flux is calculated by: $Q = -\lambda \left( \partial T / \partial y \right)_{y=0}$, where the local temperature gradient is obtained through a second-order finite difference scheme. The results reveal that the cold substrate produces the heat flux density and the temperature gradient. As freezing progresses, the decreasing unfrozen volume reduces latent heat release, thereby lowering the absolute of heat flux. From the heat flux distribution in Fig. 15, it can be seen that the magnetic field exerts little influence on the contact line between the droplet and substrate, its orientation plays a significant role in the overall heat transfer behavior. A vertical magnetic field elongates the droplet, increasing the effective thermal resistance and extending the freezing time. Conversely, a horizontal magnetic field flattens the droplet, reducing thermal resistance and consequently shortening the freezing duration. During solidification, most of the heat released by the droplet is consumed as latent heat rather than sensible heat. Therefore, variations in thermal resistance directly influence the overall freezing efficiency.

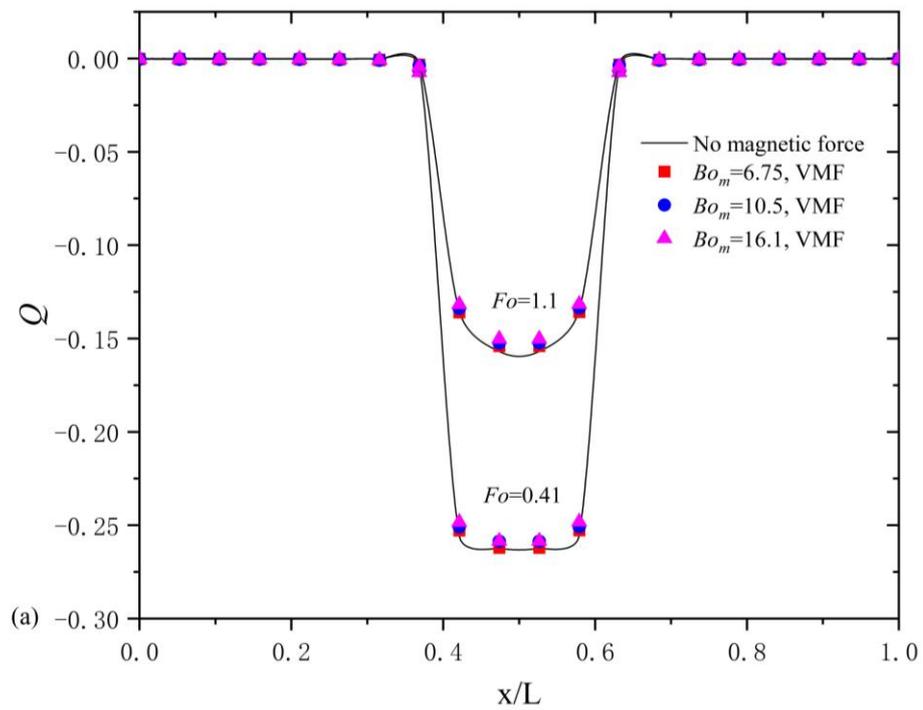

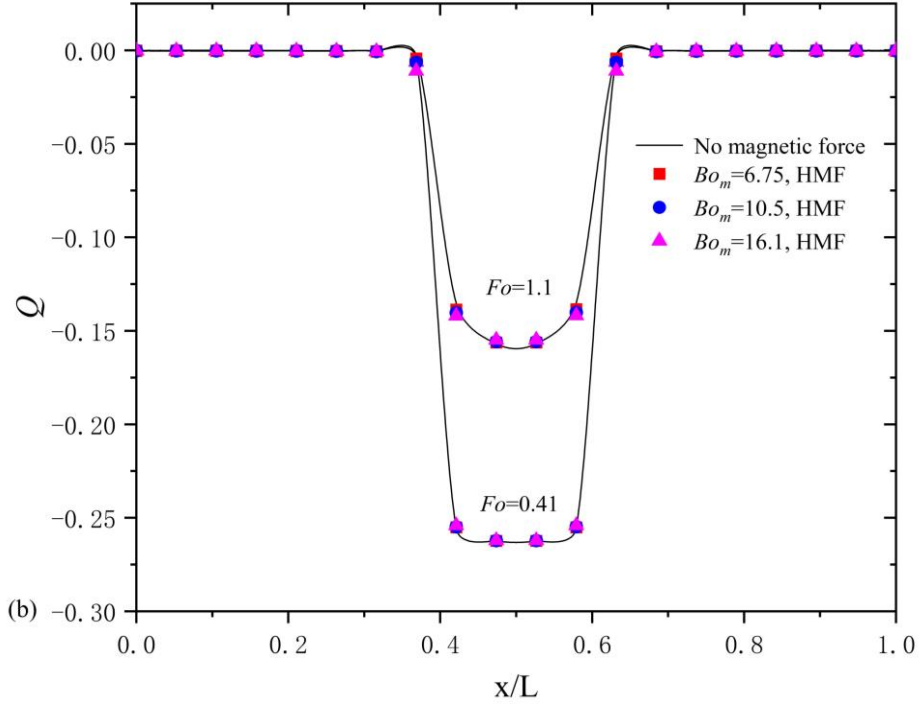

Fig. 15 Local heat flux distributions on the cold substrate at different *Fo* numbers under (a) vertical and

(b) horizontal uniform magnetic fields.

To further elucidate this effect, Fig. 16 presents the evolution of the dimensionless latent heat release rate of the droplet under different uniform magnetic field intensities, defined as:

$$Q_L = \frac{1}{A_0} \int_{\Omega_2} F_s(\mathbf{x}, t) dA, \tag{51}$$

where $A_0$ is the droplet volume and $\Omega_2$ means the region inside the droplet. The results show that under a vertical magnetic field, increasing field strength slows the release of latent heat, while under a horizontal magnetic field, a stronger field accelerates it. This contrasting behavior arises from magnetic-field-induced morphological changes, which alter the droplet's effective thermal resistance.

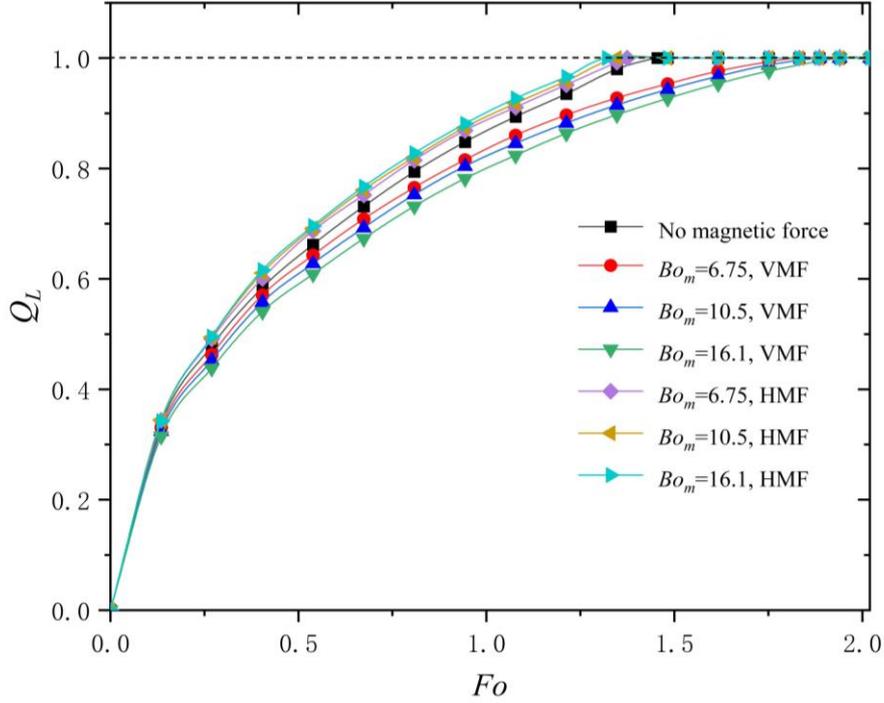

Fig. 16 Evolution of dimensionless latent heat release rate of the droplet under varying magnetic field strengths.

### 5.3 Effect of uniform magnetic field on freezing time

Finally, the effect of a uniform magnetic field on the freezing time of a ferrofluid droplet is examined. Based on Stefan's one-dimensional solidification solution, the total freezing duration can be expressed as $t_{total} = \left(H_m/2\bar{\lambda}\right)^2/\alpha$, where $H_m$ denotes the final height of the frozen droplet and $\bar{\lambda} = \sqrt{C_p\Delta T/2L}$ is the root of the Stefan problem equation. According to Zeng et al. [62], the relationship between $H_m$ and the freezing time can be further refined by incorporating the $Ste$ and $Fo$, leading to the scaling relationship $H_m/R_0 \propto \left(SteFo\right)^{0.5}$. Fig. 17 presents the scaling relationship between $H_m/R_0$ and $SteFo$. In all cases, the data points closely follow the black dashed line (slope=0.5), confirming that the freezing process of the ferrofluid droplet under a uniform magnetic field obeys the established scaling law. This result highlights a fundamental difference between uniform and non-uniform magnetic fields. As reported by Fang et al. [46], a non-uniform magnetic field induces significant droplet deformation and spatial redistribution of magnetic nanoparticles along the field gradient, thereby causing the freezing time to deviate from the theoretical scaling relation. In contrast, under a uniform magnetic field, the magnetic field gradient is zero, and magnetic nanoparticles experience no migration. Consequently, the droplet retains its geometric similarity, and the freezing process remains consistent with the established scaling law.

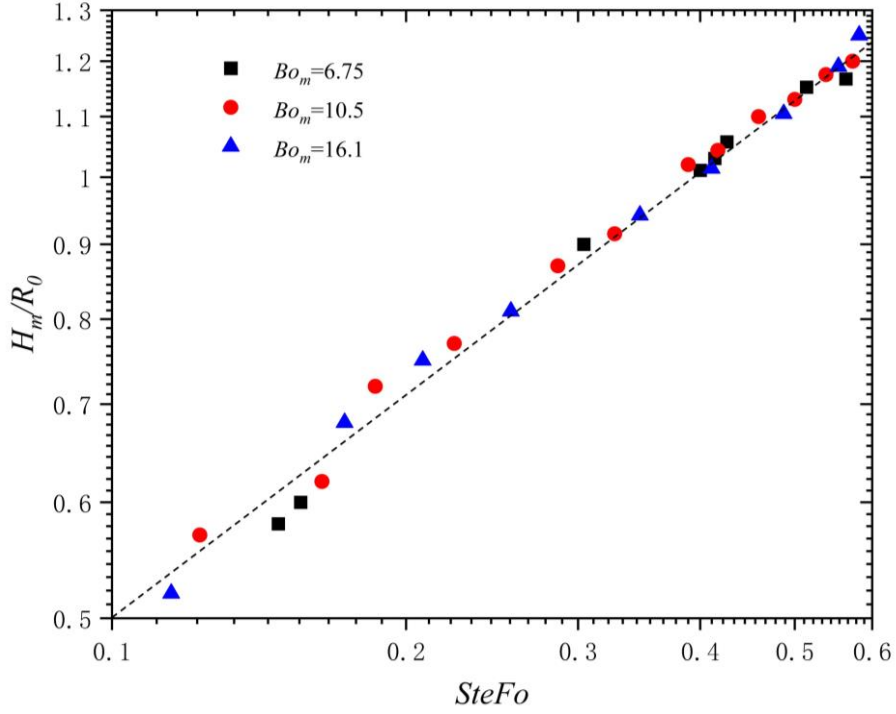

Fig. 17. Scaling relationship between $H_m/R_0$ and $SteFo$.

To further illustrate the effect of the magnetic field, Fig. 18 shows the dimensionless freezing time ($t^*=(t_{Bom}-t_{Bom=0})/t_{Bom=0}$) for the droplet with different $Ste$ numbers under varying uniform magnetic field strengths. The results indicate that a vertical uniform magnetic field prolongs the freezing time, with the duration increasing as the magnetic Bond number ($Bo_m$) becomes larger. Conversely, a horizontal uniform magnetic field shortens the freezing time proportionally with increasing $Bo_m$. As discussed in Section 4.2, a vertical magnetic field elongates the ferrofluid droplet along the field direction, increasing its height and effective thermal resistance, which leads to longer freezing durations. In contrast, a horizontal magnetic field flattens the droplet, reducing its height and thermal resistance, thereby accelerating solidification. Furthermore, as the Stefan number increases, the influence of the magnetic field on the freezing time diminishes, since higher *Ste* values correspond to shorter overall freezing durations and thus reduce the period during which the magnetic field can affect the solidification process.

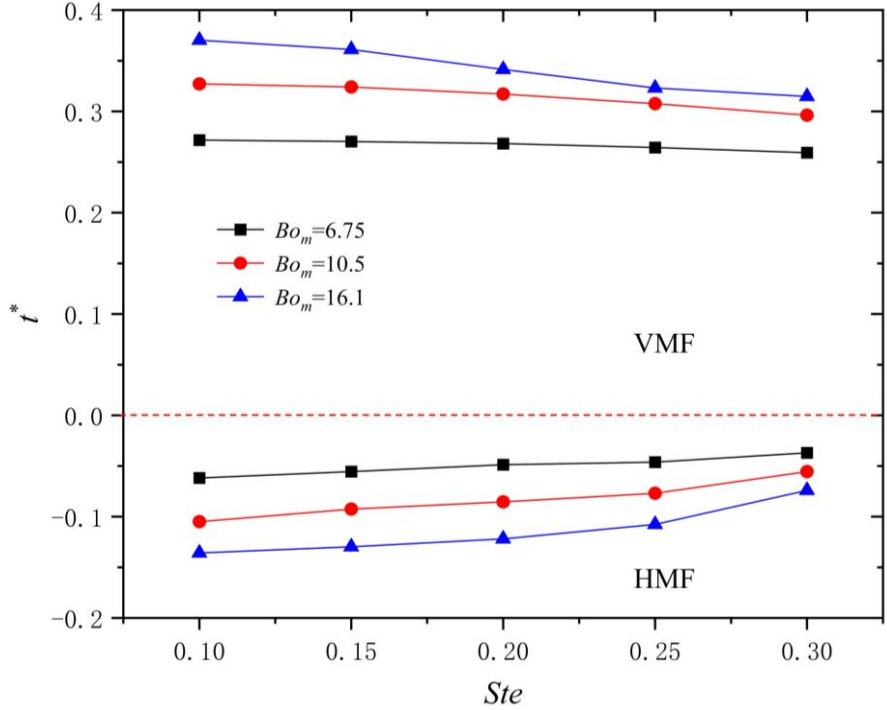

Fig. 18. Dimensionless freezing time under different $Bo_m$s and orientations.

## 6. Conclusion

In this study, an enthalpy-based lattice Boltzmann flux solver was developed to simulate the freezing dynamics of a ferrofluid droplet under a uniform magnetic field. The accuracy and reliability of the proposed solver were first verified through three benchmark tests: conductive freezing, static droplet freezing, and ferrofluid droplet deformation. Subsequently, the solver was employed to investigate the influence of a uniform magnetic field on the freezing behavior, heat transfer characteristics, and overall solidification time of the droplet.

The results demonstrate that, although the uniform magnetic field exerts only a minor influence on the contact line between the droplet and the cold substrate, it noticeably affects the overall freezing dynamics. A vertical uniform magnetic field elongates the droplet along the field direction, thereby increasing the effective thermal resistance and prolonging the freezing time. In contrast, a horizontal uniform magnetic field flattens the droplet, leading to reduced thermal resistance and a shorter freezing duration. While magnetic-field-induced convection slightly enhances internal circulation, its contribution to heat transfer remains limited due to the dominant role of thermal diffusion during solidification. Furthermore, the freezing time of the ferrofluid droplet under a uniform magnetic field follows the classical scaling law $H_m/R_0 \propto (SteFo)^{0.5}$, indicating that the magnetic field primarily affects the freezing process by modifying the droplet morphology rather than altering the intrinsic heat transfer mechanism.

**Acknowledgments**

The research is supported by the National Natural Science Foundation of China (12202191, 92271103) and the State Key Laboratory of Mechanics and Control for Aerospace Structures (Nanjing University of Aeronautics and Astronautics) (MCAS-S-0324G03).

**Appendix. Chapman-Enskog expansion analysis for the temperature field**

The Taylor series expansion of Eq. (19) can be expressed as:

$$\left(\frac{\partial}{\partial t}+\mathbf{e}_\alpha \cdot \nabla\right)h_\alpha + \frac{1}{2}\delta_t\left(\frac{\partial}{\partial t}+\mathbf{e}_\alpha \cdot \nabla\right)^2 h_\alpha + o\left(\delta_t^2\right) = -\frac{1}{\tau_h \delta_t}\left(h_\alpha - h_\alpha^{eq}\right). \quad \text{(A-1)}$$

By introducing the multi-scale expansion:

$$\begin{cases} \dfrac{\partial}{\partial t} = \varepsilon^1 \dfrac{\partial}{\partial t_1} + \varepsilon^2 \dfrac{\partial}{\partial t_2} \\ h_\alpha = h_\alpha^{(0)} + \varepsilon^1 h_\alpha^{(1)} + \varepsilon^2 h_\alpha^{(2)}, \\ \nabla_x = \varepsilon^1 \nabla_{x1} \end{cases} \quad \text{(A-2)}$$

where $\varepsilon$ is a small parameter proportional to the Knudsen number. Substituting Eq. (A-2) into (A-1) yields a hierarchy of equations at successive orders of $\varepsilon$:

$$\begin{cases} o(\varepsilon^0): h_\alpha^{(0)} = h_\alpha^{eq} \\ o(\varepsilon^1): \dfrac{\partial}{\partial t_1} h_\alpha^{(0)} + \mathbf{e}_\alpha \cdot \nabla_{x1} h_\alpha^{(0)} = -\dfrac{1}{\tau_h \delta_t} h_\alpha^{(1)} \\ o(\varepsilon^2): \dfrac{\partial}{\partial t_2} h_\alpha^{(0)} + \dfrac{\partial}{\partial t_1}\left(1-\dfrac{1}{2\tau_h}\right) h_\alpha^{(1)} + \mathbf{e}_\alpha \cdot \nabla_{x1}\left(1-\dfrac{1}{2\tau_h}\right) h_\alpha^{(1)} = -\dfrac{1}{\tau_h \delta_t} h_\alpha^{(2)} \end{cases} \quad \text{(A-3)}$$

The moments of the equilibrium distribution function satisfy:

$$\begin{cases} \sum_\alpha h_\alpha^{eq} = H_e \\ \sum_\alpha \mathbf{e}_\alpha h_\alpha^{eq} = C_p T \mathbf{u} \\ \sum_\alpha \mathbf{e}_\alpha \mathbf{e}_\alpha h_\alpha^{eq} = C_p T \mathbf{u}\mathbf{u} + C_{p,ref} c_s^2 \mathbf{I} \end{cases} \quad \text{(A-4)}$$

Using the compatibility condition, we obtain:

$$\sum_\alpha h_\alpha^{(k)} = 0, \quad (k \geq 1). \quad \text{(A-5)}$$

Substituting Eqs. (A-4) and (A-5) into Eq. (A-3) gives:

$$\begin{cases} \varepsilon^1: \dfrac{\partial H_e}{\partial t_1} + \nabla_{x1} \sum_\alpha \mathbf{e}_\alpha h_\alpha^{(0)} = 0 \\ \varepsilon^2: \dfrac{\partial H_e}{\partial t_2} + \nabla_{x1} \cdot \left[\left(1-\dfrac{1}{2\tau_h}\right)\sum_\alpha \mathbf{e}_\alpha h_\alpha^{(1)}\right] = 0 \end{cases} \quad \text{(A-6)}$$

Combining the resultant formulations at time scales $t_1$ and $t_2$ and defining $h_\alpha^{neq} = \varepsilon h_\alpha^{(1)}$, the macroscopic equation for the temperature field can be derived:

$$\dfrac{\partial H_e}{\partial t} + \nabla \cdot \left[\left(\sum_\alpha \mathbf{e}_\alpha h_\alpha^{eq}\right) + \left(1-\dfrac{1}{2\tau_h}\right)\left(\sum_\alpha \mathbf{e}_\alpha h_\alpha^{neq}\right)\right] = 0. \quad \text{(A-7)}$$

According to Eqs. (A-3) and (A-4), we can deduce:

$$-\dfrac{1}{\tau_h \delta_t} \sum_\alpha \mathbf{e}_\alpha h_\alpha^{(1)} = \nabla_{x1} \cdot \left(C_{p,ref} T c_s^2 \mathbf{I}\right) + \dfrac{\partial (C_p T \mathbf{u})}{\partial t_1} + \nabla_{x1} \cdot \left(C_p T \mathbf{u}\mathbf{u}\right). \quad \text{(A-8)}$$

Substituting Eq. (A-8) into (A-7) and neglecting the higher order terms $\dfrac{\partial (C_p T \mathbf{u})}{\partial t_1}$ and $\nabla_{x1} \cdot (C_p T \mathbf{u}\mathbf{u})$[41], Eq. (A-7) can be simplified as:

$$\frac{\partial H_e}{\partial t}+\nabla\cdot\left(C_p T\mathbf{u}\right)=\nabla\cdot\left[c_s^2\left(\tau_h-0.5\right)\delta_t C_{p,ref}\nabla T\right]. \tag{A-9}$$

The relationship between $\tau_h$ and $\lambda$ can be expressed as:

$$\tau_h=\frac{\lambda}{\rho c_s^2 \delta_t C_{p,ref}}+0.5. \tag{A-10}$$

Substituting Eq. (A-10) into (A-9) further simplifies the expression:

$$\frac{\partial H_e}{\partial t}+\nabla\cdot\left[C_p T\mathbf{u}-\frac{\lambda}{\rho}\nabla(T)\right]=0. \tag{A-11}$$

By comparing Eq. (A-11) and Eq. (A-7), the relationship between the distribution function and the macroscopic flux can be established:

$$C_p T\mathbf{u}-\frac{\lambda}{\rho}\nabla(T)=\sum_\alpha \mathbf{e}_\alpha\left(h_\alpha^{eq}+\left(1-\frac{1}{2\tau_h}\right)h_\alpha^{neq}\right). \tag{A-12}$$

By introducing $h_\alpha^*=\left[h_\alpha^{eq}+\left(1-\frac{1}{2\tau_h}\right)h_\alpha^{neq}\right]$, the above relationship simplifies to:

$$C_p T\mathbf{u}-\frac{\lambda}{\rho}\nabla(T)=\sum_\alpha \mathbf{e}_\alpha h_\alpha^*. \tag{A-13}$$

Eq. (A-13) forms the theoretical foundation for computing the flux in the total enthalpy equation.